*Sequence analysis*

# SimPlot++: a Python application for representing sequence similarity and detecting recombination

Stéphane Samson[1], Étienne Lord[2] and Vladimir Makarenkov[1*]

[1]Department of Computer Sciences, Université du Québec à Montréal, Montreal, QC, Canada

[2]Agriculture and Agri-Food Canada, Saint-Jean-sur-Richelieu, QC, Canada

*To whom correspondence should be addressed.

**Abstract**

**Motivation:** Accurate detection of sequence similarity and homologous recombination are essential parts of many evolutionary analyses.

**Results:** We have developed SimPlot++, an open-source multiplatform application implemented in Python, which can be used to produce publication quality sequence similarity plots using 63 nucleotide and 20 amino acid distance models, to detect intergenic and intragenic recombination events using $\Phi$, Max-$\chi^2$, NSS or proportion tests, and to generate and analyze interactive sequence similarity networks. SimPlot++ supports multicore data processing and provides useful distance calculability diagnostics.

**Availability:** SimPlot++ is freely available on GitHub at: https://github.com/Stephane-S/Simplot_PlusPlus, as both an executable file (for Windows) and Python scripts (for Windows/Linux/MacOS)

**Contact:** makarenkov.vladimir@uqam.ca

## 1 Introduction

The number of molecular sequences of interest to biologists and bioinformaticians stemming from microbial genomes and the environment continues to increase exponentially. Analysis of these sequences is commonly performed using different sequence similarity tools and methods for identifying homologous recombination events. Recombination is a prevalent process contributing to genetic and functional diversity of most bacteria and viruses, allowing them to overcome selective pressure and adapt to new environments (Posada & Crandall, 2001; Pérez-Losada et al., 2015). Genetic recombination, which is a key mechanism of reticulate evolution (Makarenkov and Legendre, 2000; Makarenkov et al., 2004) can be either intergenic, implying that a complete copy of a new gene is incorporated into the host genome or intragenic, implying the emergence of mosaic genes (or chimeras), formed from DNA fragments of various origins as a result of gene fusion process. Quick and accurate identification of eventual intergenic and intragenic recombination events helps improve our understanding of evolutionary history of species under study and discover functional differences induced by these events (Patiño-Galindo et al., 2021). The existing programs for detecting recombination events include TOPALi (Milne et al., 2004), RAT (Etherington et al., 2005), Armadillo 1.1 (Lord et al., 2012), T-Rec (Tsimpidis et al., 2017), and RDP4 and 5 (Martin et al., 2015 and Martin et al., 2021).

While several software for assessing and visualizing sequence similarity, using either traditional similarity plots (SimPlot by Lole et al., 1999; GenAlyzer by Choudhuri et al. 2004; Circoletto by Darzentas, 2010) or sequence similarity networks (Cytoscape by Shannon et al. 2003; genBaRcode by Thielecke et al. 2020) have been proposed, the SimPlot program remains a tool of choice of many researchers, despite the fact that it was developed in 1999 and is only distributed as a Windows 32-bit application.

In this paper, we describe SimPlot++, a reinterpretation of the original SimPlot (Lole et al., 1999) program, which comprises all its basic functionalities, including publication quality SimPlot and BootScan similarity graphs. In addition, our open-source multiplatform application implemented in Python (its Windows, Mac and Linux/Unix versions are available on our GitHub repository) offers users a large variety of nucleotide and amino acid distance models which can be used for calculating sequence similarity, detecting intergenic and intragenic recombination events, and generating and analyzing interactive sequence similarity networks. Furthermore, SimPlot++ supports multicore computation and provides helpful sequence quality (i.e. pairwise distance calculability) diagnostics, which are usually unavailable in other packages.

## 2 Methods

SimPlot++ takes advantage of a PyQt5 graphical user interface. When it is launched, the user is asked to load a multiple sequence alignment file in the FASTA, Nexus or other traditional alignment format. The loaded sequences can then be easily divided into different groups that will be used to calculate consensus sequences. The resulting consensus sequences serve as the basis for the SimPlot, BootScan, Network similarity and Recombination analyses. Once calculated, the user can inspect each consensus sequence in order to set an appropriate confidence threshold. To speed up the computations, a multi-core option has been made available to the users and the Numba just-in-time compiler was used to obtain best performance from code.

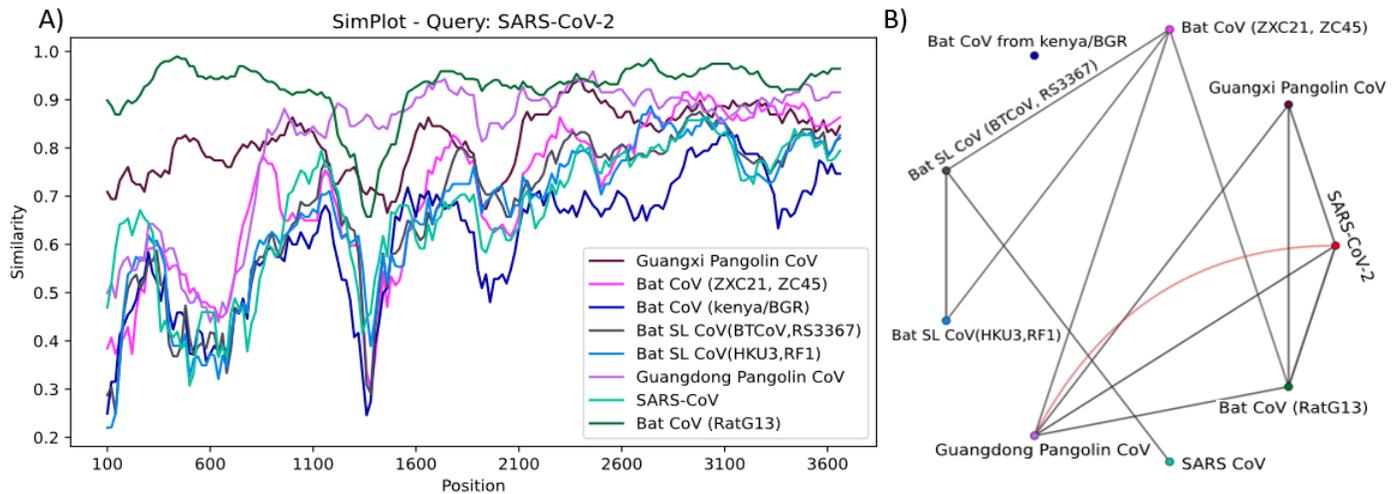

**Fig. 1. Results of the SimPlot (a) and Network (b) analyses performed using SimPlot++.** Gene S nucleotide sequences of SARS-CoV-2 (Wuhan's strain) and of 8 different coronavirus groups (consensus group sequences) including human, bat and pangolin CoV strains were used in this analysis (for data description, see Makarenkov et al., 2021). (a) SimPlot analysis was carried out using as reference the SARS-CoV-2 gene sequence; (b) Sequence similarity network depicts the most important global (black edges) and local (red edge) similarities between gene sequences. Thresholds for the global and local similarities have been set to 76% and 85%, respectively. The red edge represents a high similarity, implying a possible intragenic recombination event, detected between the SARS-CoV-2 and Guangdong Pangolin sequences in the 1220-1520nt region of gene S corresponding to its receptor-binding (RB) domain.

The basic SimPlot analysis consists of a graphical representation of similarities between a reference group and the remaining (consensus) groups using a sliding window approach (Fig. 1a). The SimPlot analysis can be conducted using 63 different nucleotide distance models and 20 different amino acid distance models. Moreover, a distance calculability diagnostic tool incorporated in our program allows the user to identify the alignment fragments for which the distance computation was impossible due to insufficient sequence similarity or high gap frequency. The BootScan (Salminen et al., 1995) and FindSite tools (Robertson et al., 1995) can be used, respectively, for the identification and mapping of recombinant sequence regions and the detection of informative sites.

Sequence similarities can also be conveniently visualized using network-like structures (Atkinson et al., 2009; Thielecke et al., 2020; Xing et al., 2020). In SimPlot++, the results of SimPlot analysis can be represented as a sequence similarity network (Fig. 1b) in which the nodes represent the sequences, and the edges represent global and/or local similarities between them. Different filters allowing the user to visualize the network by means of the selected global and local similarity thresholds are available. The integrated Matplotlib tool bar allows for easier exploration and customization of the produced similarity and network plots (Fig. 1).

Finally, SimPlot++ can be also used to detect intergenic recombination events using the well-known $\Phi$, Max-$\chi^2$ and NSS tests (Bruen et al., 2006), and a quick Proportion test. Moreover, $\Phi$-test, carried out in profile mode, and our Proportion test can be used to detect intragenic recombination as well.

## Funding

This work has been supported by Fonds Québécois de la Recherche sur la Nature et les Technologies [grant 173878] and Natural Sciences and Engineering Research Council of Canada [grant 249644].